# Reaction paths of alane dissociation on the Si(001) surface

Richard Smith1;2;3, David R. Bowler1;2;3;4

1. London Centre for Nanotechnology, UCL, 17-19 Gordon St, London WC1H 0AH, UK

2. Department of Physics & Astronomy, UCL, Gower St, London WC1E 6BT, UK

3. Thomas Young Centre, UCL, Gower St, London WC1E 6BT, UK

4. International Centre for Materials Nanoarchitectonics (MANA), National Institute for Materials Science (NIMS), 1-1 Namiki, Tsukuba, Ibaraki 305-0044, JAPAN

E-mail: david.bowler@ucl.ac.uk

**Abstract**

Building on our earlier study, we examine the kinetic barriers to decomposition of alane, AlH3, on the Si(001) surface, using the nudged elastic band (NEB) approach within DFT. We find that the initial decomposition to AlH with two H atoms on the surface proceeds without a significant barrier. There are several pathways available to lose the final hydrogen, though these present barriers of up to 1 eV. Incorporation is more challenging, with the initial structures less stable in several cases than the starting structures, just as was found for phosphorus. We identify a stable route for Al incorporation following selective surface hydrogen desorption (e.g. by STM tip). The overall process parallels PH3, and indicates that atomically precise acceptor doping should be possible.

# 1 Introduction

## 1.1 Background

Recently, phosphine (PH3) has emerged as the favoured precursor for precision donor doping of silicon at the atomic scale, using hydrogen-resist patterned atomic layer epitaxy (PALE) [1,2,3]. The timely availability of acceptor dopants via PALE would increase the variety of devices which could be fabricated, e.g. the tunnel FET where accurate dopant placement is a large advantage. Another application for p-type dopants is to improve n-type devices by increasing the effective barrier around active elements. For example, the necessary interaction between qubit devices defines a maximum inter-qubit spacing of around 20 nm [4]. However, each qubit also needs control and sensing electrodes affecting only one qubit. By introducing acceptor dopants into the background between the electrodes their potential wells could be more closely confined, minimizing crosstalk.

Given the widespread use of boron as an acceptor dopant in LSI devices, it is natural to suppose that diborane ($B_2H_6$) will assume a complementary role to phosphine at the nanoscale. However, diborane is not an ideal dopant choice, as it has a low sticking coefficient and cannot readily deposit a *single* boron atom, as would be required by the PALE process [5,6]. It is therefore important to consider other possibilities.

In a previous communication [7], we proposed alane (aluminium hydride $AlH_3$) as a suitable precursor for Al deposition as an acceptor dopant. Although difficult to synthesize, an energetic analysis implies that it will adsorb and dissociate on the Si (100) surface, yielding Al bonded in dimer-bridging modes. We assumed that the H ligands of alane would detach sequentially, re-adsorbing in the immediate vicinity as should occur in a PALE process. We also investigated surface configurations having an incorporated Al atom. We found that many were less stable than the bridging configurations available as starting points (this relative stability is also seen with P incorporation [8]). In the absence of kinetic barriers, one might expect reversal of the incorporation reaction, unless another forward pathway leading to a configuration with better stability is available.

We now resolve these results by providing a kinetic analysis of the available dissociation and incorporation pathways of alane on the Si(100) surface. Our DFT calculations yield the activation energies and hence the expected reaction rates of each step under PALE process conditions. We illustrate maximum-energy (transition state) configurations in diagrammatic form and suggest an incorporation procedure based on STM removal of dissociated H prior to incorporation.

These findings, taken together with our earlier results, provide a testable route to Al incorporation in the Si(100) surface and will support experimental work in this area.

## 2  Methods

### 2.1  Terminology

We continue with the configuration naming scheme introduced earlier and summarized at figure 1. A pathway segment is defined by a starting and ending configuration e.g. A301-B202 is a path between the initial dimer-end adsorption configuration A301 and the bridged-dimer first-dissociation configuration B201. Fully dissociated configurations (having a bare Al adatom) have zero as the second character of the name e.g. B002, D004. Incorporation configurations are categorized as end-bridge (because they give rise to an Si adatom in that configuration) and are numbered sequentially from 50 upwards, i.e. D050, D051 and so on.

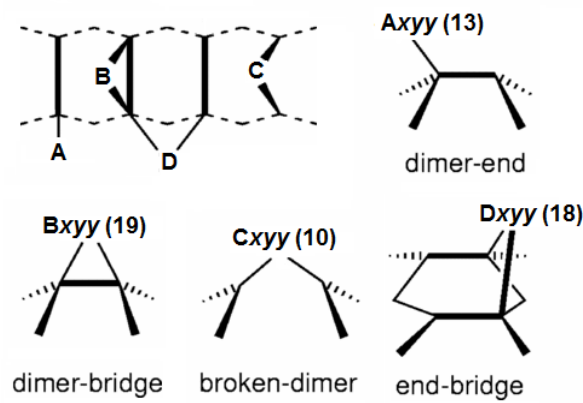

*Figure 1* Perspective views of adsorption sites of AlH$_3$ on the Si(100) surface, reproduced from [7].  Adatom A binds at a *dimer-end* position of a surface dimer; B binds to two Si atoms on the same dimer in the *dimer-bridge* position, leaving the dimer intact; *broken-dimer* position C is like B, but breaks the dimer and D binds to Si atoms on two adjacent dimers in the *end-bridge* position.  Dissociation is modelled by removing an H from the adatom and placing it nearby.  This creates a new surface configuration identified by appending a number *xyy* where x indicates the number of H atoms remaining bonded to Al, i.e. *x=3* represents the initial adsorption, *x=0* indicates a fully dehydrogenated Al atom.  yy is an enumerator.  The respective number of identified structures appears in parentheses.

## 2.2   Computational details

All calculations used density functional theory [9], as implemented in the Vienna Ab-initio Simulation Package (VASP versions 5.3.1/5.4.1) [10] with the Perdew-Burke-Enzerhof (PBE) generalised gradient approximation (GGA) exchange-correlation functional [11]. The VASP projector-augmented-wave [11, 12] (PAW) potentials for aluminium, silicon and hydrogen were used.  These potentials describe both core and valence electrons and the files (POTCAR) were dated 4/5th January 2001 and 15th June 2001, respectively.  We used a 400 eV energy cut-off.  This value is required for proper operation of the aluminium PAW pseudopotential.

The Si(100) surface was modelled on a slab of eight Si layers with a c(4x2) surface cell reconstruction, separated by a 12 Å vacuum gap.  The 15.36 Å x 15.36 Å surface accommodates two dimer rows of buckled dimers (four in each row) at approximately 18° to the surface plane.  The relatively large surface supports adsorption configurations spanning adjacent dimer rows. The bottom layer of Si atoms was left in bulk-like positions, terminated with pairs of hydrogen atoms, and fixed.  The experimental bulk Si lattice parameter (5.431 Å) was used and is within 1% of the PBE lattice constant.  The Brillouin zone was sampled with a 3x3x1 Monkhorst-Pack mesh [13].

Transition state search was performed using the climbing image nudged elastic band method, as implemented in the VASP Transition State Tools [14] (VTST) code.  This method is known to yield accurate barrier energy values [16].  The calculation was initialized using by six intermediate images on each path segment, obtained by linear interpolation from the minimum-energy end points.  The convergence criterion for atom forces was set to 0.05 eV/Å and that for total energy to $10^{-5}$ eV, and the maximum allowable number of structural optimizations was set to 200. We used the FIRE algorithm [17] to optimize the intermediate structures.

## 2.3 Nudged elastic band method

NEB methods [14] provide a discrete representation of the minimum energy path (MEP) between the initial and final states of a chemical transition such as dissociative adsorption. Each point on the MEP corresponds to an image of the system, i.e. a set of atomic coordinates. A path is an MEP only if the force (corresponding to the energy gradient) on each image is tangential to the path. The maxima on the MEP are saddle points on the potential energy surface. The rate constant for the transition through each saddle point is derived statistically from its energy value using transition state theory and the Arrhenius relation.

An initial approximation to the MEP is obtained by the linear interpolation of images between the initial and final states, with the images connected with springs to form a band. A band with $N + 1$ images can be denoted by $[\hat{R}_0, \hat{R}_1, \hat{R}_2, \ldots, \hat{R}_N]$ where $\hat{R}_0$ and $\hat{R}_N$, which are fixed, correspond to the initial and final states. Then the $N - 1$ intermediate images are adjusted using a projection scheme, characteristic of NEB methods. The force acting on the image is taken to be the sum of the projections of the true (potential-derived) force perpendicular to the local tangent and the spring force along the local tangent:

$$\hat{F}_i = \hat{F}_i^s\big|_\parallel - \nabla E(\hat{R}_i)\big|_\perp$$

where the perpendicular component of the true force is given by subtracting out the tangential component

$$\nabla E(\hat{R}_i)\big|_\perp = \nabla E(\hat{R}_i) - \left(|\nabla E(\hat{R}_i)|.\hat{\tau}_i\right)\hat{\tau}_i$$

and $E$ is the energy of the system and $\hat{\tau}_i$ the normalized local tangent at image $i$. The spring force is:

$$\hat{F}_i^s\big|_\parallel = k\left(|\hat{R}_{i+1} - \hat{R}_i| - |\hat{R}_i - \hat{R}_{i-1}|\right)\hat{\tau}_i.$$

The images are all instantaneously moved along the force vectors $\hat{F}_i$ using a projected velocity Verlet algorithm [14]. The images converge on the MEP ($\hat{F}_i \to 0$) with equal spacing if the spring constant is uniform, but will not necessarily coincide with any saddle points. The actual saddle point energy must be obtained by interpolation. The climbing-image NEB (Cl-NEB) method [15] is a refinement that places one of the images precisely on the highest saddle point. After a few iterations of the regular method, the image with the highest energy $i_{max}$ is selected. The force on this one image is now given by:

$$\hat{F}_{i_{max}} = -\nabla E \hat{R}_{i_{max}} + 2\nabla E(\hat{R}_i)\big|_\parallel$$

$$= -\nabla E \hat{R}_{i_{max}} + 2\left(|\nabla E(\hat{R}_{i_{max}})|.\hat{\tau}_i\right)\hat{\tau}_i.$$

This is the true force with the component along the local path inverted, and the image is unaffected by the spring forces. The climbing image is seen to move up the potential energy surface along the path and downwards perpendicular to the path, but converges on the saddle point.

## 2.4 Convergence considerations

As noted, potential pathways were chosen based on step-wise alane dissociation through MEPs having end-points known to be favoured energetically. Even so, many apparently plausible scenarios failed to yield a converged result. In the NEB context, convergence means that the total force (with the perpendicular component of the potential gradient projected out and the parallel component of the spring force added in) is less than a global limiting value for each intermediate image.

Images are constructed by linear interpolation and resulting atomic trajectories may be unrealistic, especially when modelling an incorporation reaction. When inter-atomic distances become small large repulsive forces are generated, and the NEB algorithm can sample high-energy regions of the potential energy surface without ever satisfying the convergence conditions. For this reason, the conditions are moderated when compared with a structural energy minimization (see section 2.2).

Difficulties also arise when the dissociating H in an interpolated image approaches that in another known configuration. Since the latter is located at a local energy minimum subsequent movement will be confined. The resulting MEP must then traverse steep energy gradients as the spurious Si-H bond is broken, and the calculation may again terminate without result. In most such cases the intermediate configuration appeared as an end-point on other MEPs, so no further action was needed. In others, the intermediate configuration showed a small gain in stability, attributed to the shallow valleys in the energy surface associated with rotational movement of the H ligands. In these cases, the intermediate configuration was re-optimized and replaced an original survey point.

## 2.5 Activation energy and reaction rate

The energy difference between the initial configuration and the highest saddle point on an MEP is the activation energy $E_A$ which is related to the reaction rate $k$ by the Arrhenius equation:

$$k = \nu e^{-E_A/k_B T}$$

where $\nu$ is the attempt frequency, $k_B$ the Boltzmann constant and $T$ the prevailing temperature. This simple formulation arises from transition state theory (TST [18]) and assumes the reaction rate is sufficiently slow for a Boltzmann energy distribution to be established in the reactants and neglects quantum effects such as zero-point energy and tunnelling. Since the rate is dominated by the exponent term it is common to use an estimate of $\nu$ (e.g. $\nu = 10^{12} - 10^{14} \, s^{-1}$) rather than attempt to calculate a more precise value.

In table 1 below we show the activation energies yielding a transition probability of one over various timescales at PALE process temperatures. These are estimates based on the Arrhenius reaction rate using the boundary values of $\nu$ given above, and may assist the reader in determining the feasibility of the reactions to be presented later.

| PALE process temperature (K) | Activation energies (eV) | | |
|---|---|---|---|
| | 1 s | 1 m | 1 h |
| 150 | 0.36-0.42 | 0.41-0.47 | 0.46-0.55 |
| 200 | 0.48-0.56 | 0.55-0.63 | 0.62-0.70 |
| 250 | 0.60-0.69 | 0.68-0.78 | 0.77-0.87 |
| 300 | 0.71-0.83 | 0.82-0.94 | 0.93-1.05 |
| 350 | 0.83-0.97 | 0.96-1.10 | 1.08-1.22 |
| 400 | 0.95-1.11 | 1.09-1.25 | 1.23-1.39 |
| 450 | 1.07-1.25 | 1.23-1.41 | 1.39-1.57 |
| 500 | 1.19-1.39 | 1.37-1.57 | 1.54-1.74 |
| 550 | 1.31-1.53 | 1.50-1.72 | 1.70-1.92 |

*Table 1* Calculated activation energies (from the TST/Arrhenius equation with $\nu = 10^{12} - 10^{14}\ s^{-1}$) attainable for activation of diffusion and incorporation events expected to occur within periods of 1 second, 1 minute and 1 hour, at various temperatures (e.g. a reaction with $E_A$ = 1.00 eV would be activated within 1 hour at 300 K). At temperatures greater than 550K the H passivation layer, essential to the PALE process, becomes mobile and ultimately desorbs from the substrate.

## 3 Results and discussion

### 3.1 *Reaction pathways*

In our earlier work [7], we evaluated approximately 70 surface configurations at various stages of dissociation. Although we expected any minimum energy pathway to traverse the more stable configurations at each stage we nevertheless examined many sterically feasible segments involving end configurations of lower stability. A segment was considered feasible if it possessed an unobstructed H migration route, but did not call for re-diffusion of previously dissociated H and the high energy levels needed to break an Si-H bond. Under these constraints approximately 110 feasible segments were available, of which 40 yielded converged results in the MEP calculation. We considered a calculation non-converging when it exceeded the nudging algorithm's iteration limit, as discussed in section 2.4 above. This limit (and the number of intermediate images) were chosen pragmatically bearing in mind the relatively large number of computationally expensive calculations undertaken. We repeated some non-converging calculations using altered intermediate locations and relaxed convergence conditions, but without additional result.

The 40 MEP segments can be combined to give 14 pathways from initial adsorption to full dissociation, with an Al adatom and three surface H. These fall into two groups of seven, differing only in their initial configurations, i.e. the A301 dimer-end configuration and the D301 end-bridge configuration. Earlier [7], we had concluded these configurations were similar in character, differing only by a correlated rotation of the three H ligands, and this is consistent with the present NEB findings. We therefore limit our discussion to the pathways that start with the A301 configuration, focussing on those that play out on adjacent dimers in a single row.

Figure 2 compares the stabilities and activation energies of the seven pathways, which include the final incorporation MEPs. The data points are summarized at table 2. VASP configuration files, including those starting on the D301 configuration, are available on figshare [19].

Reaction paths of alane dissociation on the Si(100) surface

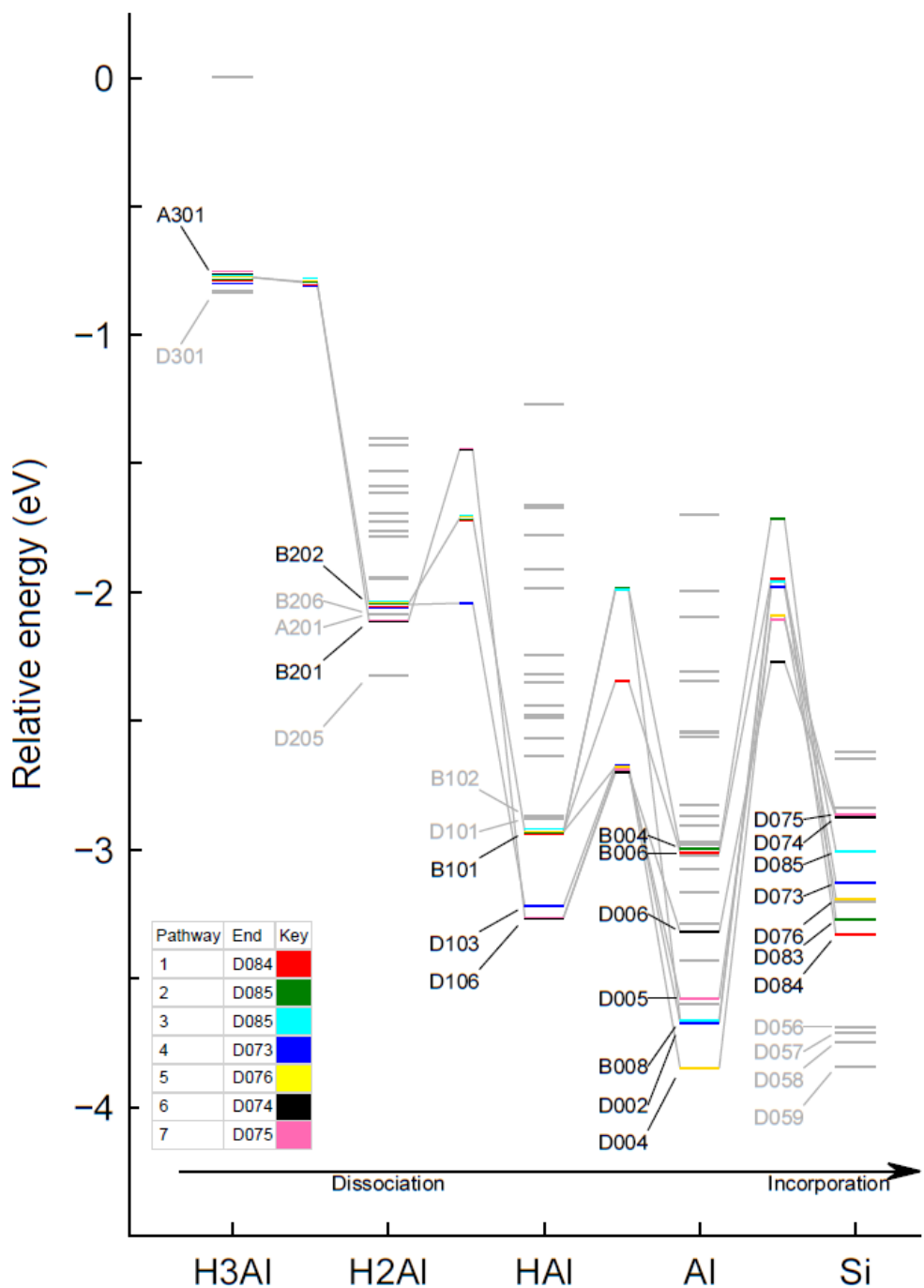

*Figure 2* Seven continuous dissociation and incorporation pathways of alane, on Si(100). Vertical columns of bars show relative configuration energies (eV) at each stage. Energies are relative to the sum of bare surface and free alane energies. A pathway is indicated by a succession of bars of the same colour, e.g. the pathway 5 through configurations *A301-B202-B101-D004-D076* is coloured yellow. Except for D301 (see text), greyed-out configurations did not lie on any pathway. The intervening columns of shorter bars indicate the calculated relative transition state energies, i.e. the highest saddle-point energy along the MEP between adjacent configurations. The data points appear in table 2 below.

An ideal pathway would pass through successive low-energy configurations, gain stability at each stage, and terminate in incorporation. In practice, the energetics favour end-bridge configurations as dissociation proceeds [7], with some intermediate configurations of this kind acquiring greater stabilities than succeeding fully-dissociated or incorporated scenarios which are our real interest. Any kinetic barrier should be surmountable at temperatures not impairing PALE passivation through H diffusion, say 450-500K. We can also expect incorporation to be hindered kinetically by the breaking of surface bonds. If a pathway fails to yield a progressive gain in stability, then (in the absence any other possibility) its reactions will tend to reverse, and the sequence terminate on the lowest energy configuration which is both stable and kinetically accessible at the prevailing temperature.

In table 2 we show the energy changes and activation energies for each dissociation and incorporation MEP, for each pathway. The remainder of our discussion is structured as follows: in sections 3.2 and 3.3 we discuss the first and second dissociations, where the relative absence of kinetic barriers characterizes all pathways. In section 3.4 we describe two pathways (1 and 2 in table 2) that, while not involving the lowest-energy configurations, are nevertheless kinetically and thermodynamically feasible and appear to terminate in stable incorporation. In section 3.5 we discuss pathways 3, 4 and 5 where low-energy fully-dissociated configurations are sampled but do not lead to stable incorporation. Pathways 6 and 7 in table 2 involve configurations occupying two surface dimer rows and are of lesser interest and not discussed, but the calculation results are available elsewhere [19]. Finally, in section 3.6 we describe an incorporation scenario involving surface migration of the ejected Si adatom.

| Pathway | MEPs $E_A/\Delta E_{cumulative}$ (eV) | | | |
|---|---|---|---|---|
| 1 | A301-B202 | B202-B101 | B101-B006 | B006-D084 |
|   | 0.00/-2.05 | 0.01/-2.93 | 0.58/-3.01 | 1.06/-3.33 |
| 2 | A301-B202 | B202-B101 | B101-B004 | B004-D083 |
|   | 0.00/-2.05 | 0.01/-2.93 | 0.95/-2.97 | 1.26/-3.27 |
| 3 | A301-B202 | B202-B101 | B101-B008 | B008-D085 |
|   | 0.00/-2.05 | 0.01/-2.93 | 0.24/-3.67 | 1.40/-3.01 |
| 4 | A301-B202 | B202-D103 | D103-D002 | D002-D073 |
|   | 0.00/-2.05 | 0.34/-2.29 | 0.37/-3.67 | 1.56/-3.13 |
| 5 | A301-B202 | B202-B101 | B101-D004 | D004-D076 |
|   | 0.00/-2.05 | 0.01/-2.93 | 0.94/-3.85 | 1.87/-3.20 |
| 6 | A301-B201 | B201-D106 | D106-D006 | D006-D074 |
|   | 0.00/-2.11 | 0.67/-3.33 | 0.59/-3.32 | 1.23/-2.87 |
| 7 | A301-B201 | B201-D106 | D106-D005 | D005-D075 |
|   | 0.00/-2.11 | 0.67/-3.33 | 0.60/-3.78 | 1.63/-2.86 |

*Table 2* Pathway number, per-MEP activation energy $E_A$ and relative cumulative energy change $\Delta E_{cumulative}$ for the seven dissociation and incorporation pathways shown at figure 2. Activation energies are derived from a six-image VASP Cl-NEB calculation. A lowering of relative energy (i.e. a negative energy change) indicates a gain in stability, and vice versa.

## 3.2 First dissociation: $AlH_3 \rightarrow AlH_2+H$

There are two MEPs involved in the first dissociation, resolving to the dimer-bridge configurations B201 (-2.11 eV, figure 3(a)) and B202 (-2.05 eV, figure 3(b)) respectively, differing only in the ultimate location of the dissociated H ligand. No MEP was found terminating on the low-energy configurations B206 and D205, which we assume are sterically inaccessible. An MEP to configuration A201 was discovered but gave rise to a relatively unstable configuration at the second dissociation (D109, -2.63 eV) from which no onward MEP was found.

The MEPs show similar stability gains (1.34 eV and 1.27 eV respectively) and are characterized by the absence of kinetic barriers, indicating that dissocation should occur immediately after the initial adsorption and proceed independently of the ambient temperature. The diagrams suggest a shallow PES basin in the vicinity of the adsorbing Si atom.

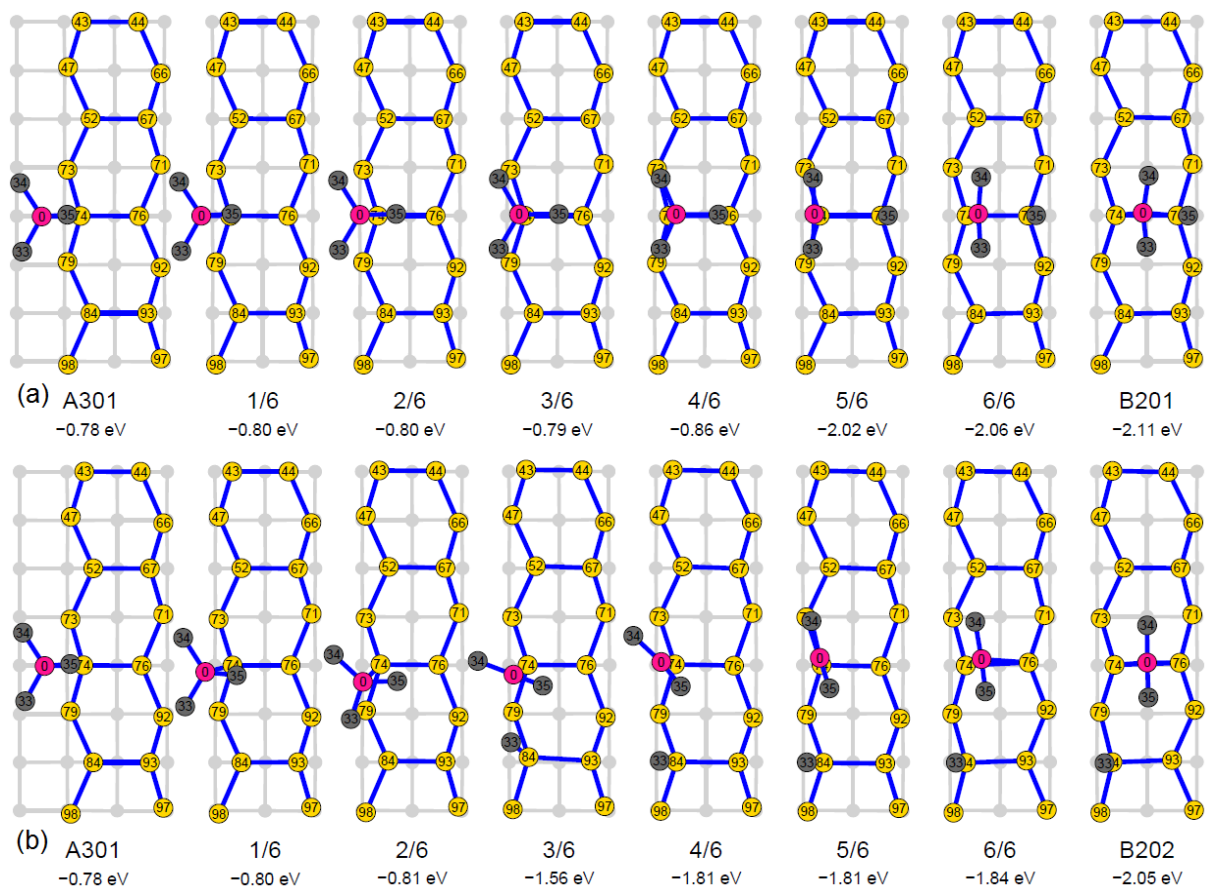

*Figure 3* Representations of two MEPs of the first dissociation of alane adsorbed on Si(100), corresponding to segments A301-B201(a) and A301-B202(b). The reaction proceeds from left to right through intermediate points 1/6, 2/6 etc. In both cases the drop in relative energies indicates the absence of kinetic barriers. Images are derived from a VASP Cl-NEB calculation and each represents a single row of alternately buckled Si dimers, viewed from above, with Si, Al and H atoms coloured yellow, pink, and grey respectively. The atom numbers are zero-based indices into the originating VASP coordinate files.

## 3.3   Second dissociation: $AlH_2+H \rightarrow AlH+2H$

The second dissociation can proceed through three MEPs, terminating in the dimer-bridge configuration B101 (-2.93 eV, figure 4(a)) and the end-bridge configurations D103 and D106 (-3.07 eV, -3.33 eV figures 4(b), (c) respectively).  However, whereas the transition to configuration B101 proceeds without significant kinetic barrier, those to the end-bridge configurations encounter barriers of 0.34 eV and 0.67 eV respectively.  The incomplete pathway to configuration D109 (-2.63 eV) mentioned above also encounters a barrier of 0.60 eV.  The TST/Arrhenius relation (section 2.5 above) indicates that even the lowest of these barriers would slow the second dissociation rate by a factor of $10^6$ compared to the barrier-free rate.  Given that the adsorbed fragment samples all directions equally we may assume reaction kinetics will dominate to ensure the onward pathways from B101 are sampled ahead of the others, possessing slightly increased stabilities.

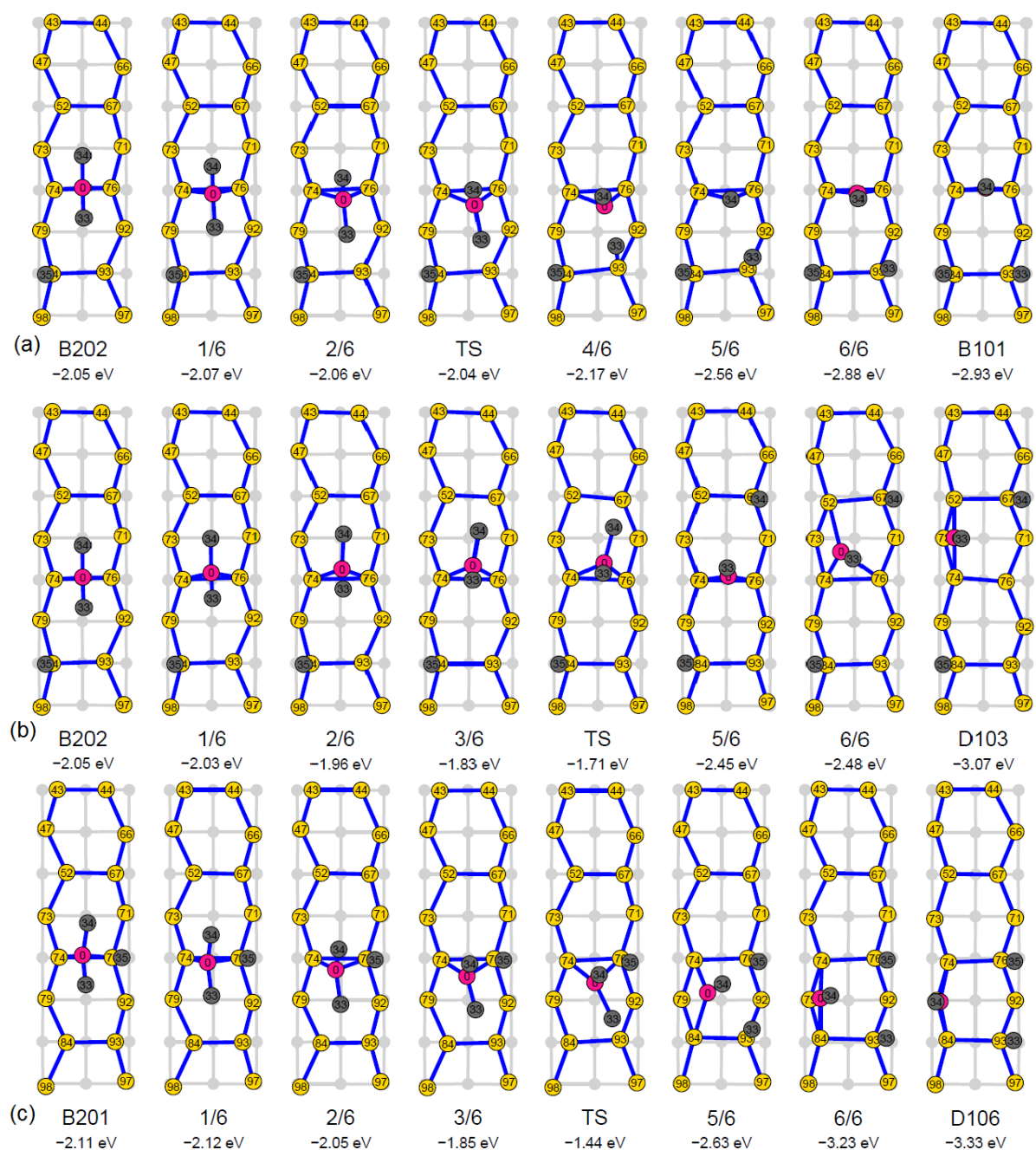

*Figure 4* Representations of three MEPs of the second dissociation of alane adsorbed on Si(100), corresponding to segments B202-B101 (a), B202-D103 (b) and B201-D106 (c). MEP (a) shows an insignificant forward barrier of 0.02 eV at image 3, whereas MEPs (b) and (c) have barriers of 0.34 eV and 0.67 eV both at image 4. Image numbering and colouring convention as for figure 3. In MEP (a), Al atom 0 is obscured by H ligand 34 in the images labelled '5/6' and 'B101'.

### 3.4 Pathways 1 & 2: stabilization on incorporation

These pathways are characterized by incorporation MEPs that show increased stability (see figure 5(a), (b)). The lowering of energy (by 0.32 eV and 0.30 eV respectively) stabilizes the incorporation through a corresponding increase in the energy barrier to reversal. The forward energy barriers (1.06 eV and 1.26 eV) are surmountable within the constraints of the PALE process and so we might consider conditioning the PALE environment to favour these pathways while frustrating others involving configuration B101. This would be feasible using an automated STM, where complex operational sequences can occur under program control. For example, on pathway 2 the dissociation plays out on just two adjacent dimers, i.e. atoms (74, 76) and (84, 93) in the figures. If these were initially de-passivated an ambient temperature of ≈ 350 K would allow the exclusive evolution of the fully-dissociated configuration B004. Then, after de-passivating a third dimer and adjusting the temperature to ≈ 450 K, incorporation in configuration D083 should follow.

Unfortunately, this scenario is unrealistic since the Al atom in configuration B004 will migrate to the lower energy end-bridge configuration D004 (seen on pathway 5, Figure 6) as soon as the third dimer is cleared. This Al migration MEP B004-D004 does not exhibit a significant activation energy and will be sampled before incorporation into D083, which presents a relatively large energy barrier. Incorporation on pathway 1 is frustrated similarly [19].

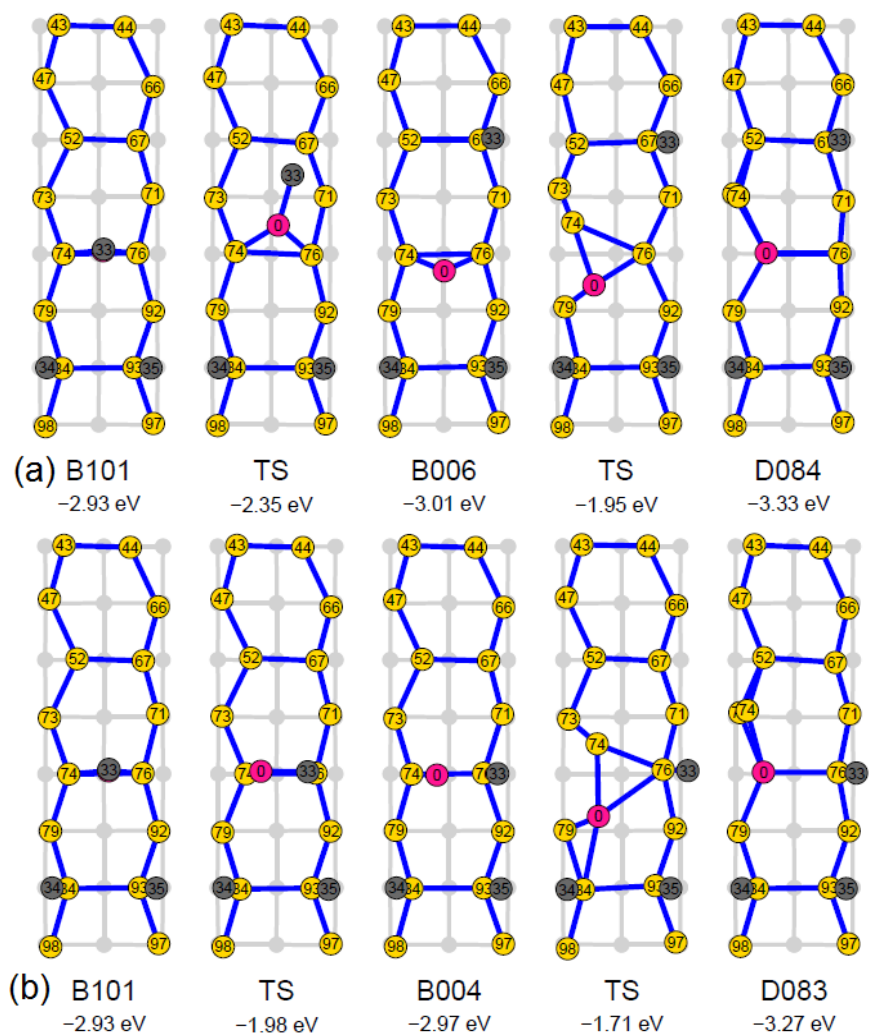

*Figure 5* MEP representations of pathways 1 (a) & 2 (b) of the third dissociation of alane adsorbed on Si(100), followed by stability-enhancing incorporation steps. On pathway 1, incorporation (B006-D084) imposes an activation energy barrier of 1.06 eV and results in a stabilization of 0.32 eV. For pathway 2, the barrier to incorporation (B004-D083) is 1.26 eV and the stabilization is 0.30 eV. Even so, these pathways fail to yield stable incorporation (see text). Image derivation and colouring convention as for figure 3. In the images of configuration B101, Al atom 0 is obscured by H ligand 33.

### 3.5 Pathways 3, 4 & 5: metastable incorporation

Figure 6 shows the third dissociation and incorporation stages of pathways characterized by destabilization on incorporation. Such destabilization (> 0.50 eV) imply that the end configurations (respectively D085, D073 and D076) are metastable and that the incorporation reactions would rapidly reverse, unless onward routes yielding sufficiently lowered energies were available. These reactions all require elevated temperatures and those on pathways 4 and 5 are not feasible under the temperature constraint of the PALE process.

At room temperatures, the kinetic barriers to incorporation are not surmountable and the fully-dissociated configurations of pathways 3 (B008) and 4 (D005) should be visible by STM inspection. The effective (considering adsorption and dissociation only) activation energies are respectively 0.24 eV and 0.37 eV. Both configurations have the same stability of -3.37 eV, but B008 is kinetically favoured and more likely to occur by a factor $\approx 10^2$. Although the fully-dissociated configuration D004 on pathway 5 is the most stable encountered, its activation energy of 0.94 eV means it is even less likely to be observed at room temperature (by a factor $\approx 10^{12}$).

If the expected evolution to configuration B008 is found to occur in practice, we could then consider some selective de-passivation that would enable stable incorporation by allowing migration of the ejected Si adatom to a new location having suitably low energy. We describe a possible scenario in the following section.

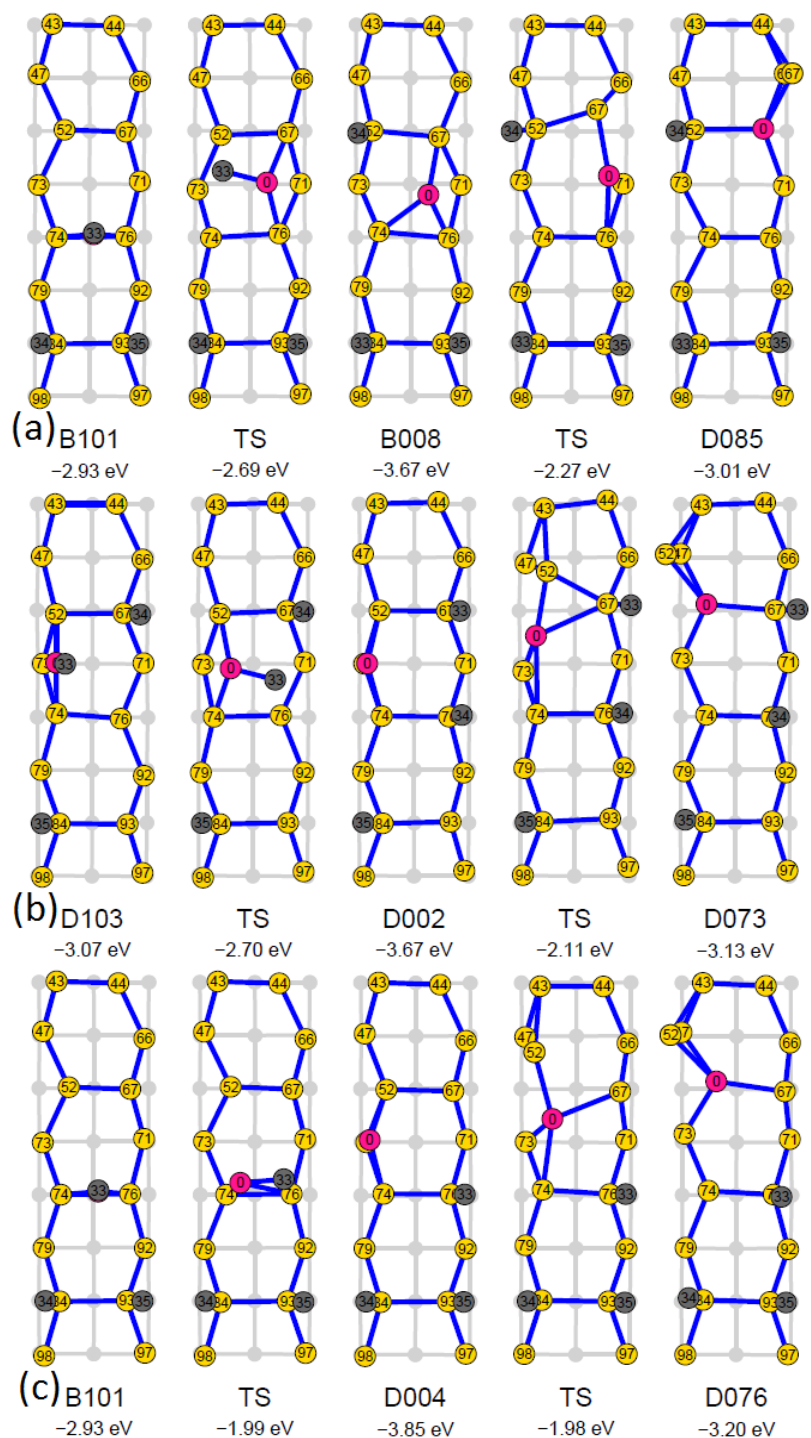

*Figure 6* MEP representations of pathways 3 (a), 4 (b) & 5 (c) for the third dissociation of alane adsorbed on Si(100), followed by incorporation. A stability loss of at least 0.50 eV in the incorporation segments B008-D085, D002-D073 and D004-D076 indicate the end configurations are metastable. The activation energies for incorporation on pathways 4 and 5 (1.56 eV, 1.87 eV) are unachievable at PALE process temperatures. At room temperatures, fully-dissociated configuration B008 on pathway 3 is likely to predominate (see text). Image derivation and colouring convention as for figure 3.

## 3.6 Post-incorporation Si migration

In our structural survey [7], we discovered a group of incorporation configurations (figure 7, D056-8) with stabilities comparable with our low-energy fully-dissociated configurations (figure 5(a) B008, (b) D002, (c) D004). However, all these incorporation configurations possess a 3-coordinated surface Al atom with an ejected Si adatom bridging a pair of Si-Si dimers. This organization will not evolve naturally during a low-temperature incorporation, which necessarily results in a 4-coordinated Al with the Si adatom bridging the Al-Si heterodimer and adjacent Si-Si dimer. The 4-coordinated configurations are seen to be less stable than those having 3-coordinated Al, which (in the PALE environment) could only be attained by exposing additional dimers so that the Si adatom could migrate away from the Al end of the heterodimer.

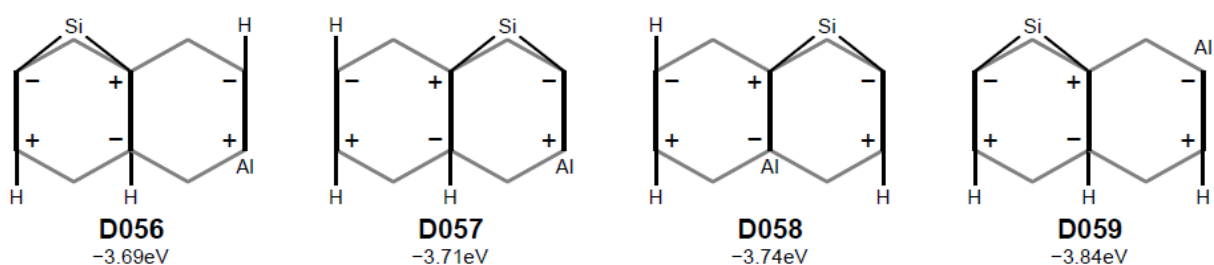

*Figure 7* Low-energy incorporation configurations discovered in the structural survey [7], and also appearing on figure 2. These are more stable than the fully dissociated configurations on pathways 3 and 4 while D059 has a stability comparable with that of D004 on pathway 5 (see also figure 2). This is due to the more favourable 3-coordination of the Al atom seen here.

We speculated that migration of the ejected Si adatom might lower the energies of our metastable incorporation configurations, thereby yielding an unconditionally stable configuration. We focussed on migration away from configuration D085 on pathway 3, seeking configurations with relative energies significantly less than its predecessor B008 (-3.67 eV), but were unsuccessful. We did find examples with lowered energies, but none sufficiently low to stabilize the incorporation.

Equivalently, we sought to destabilize the B008 configuration by removing its adsorbed H atoms (H34, H35 and H36 see figure 6(a)). The new configuration, B021, was structurally optimized and showed Al-Si bond lengths increased by ≈ 1% compared to B008, with the adsorbate coordination remaining unchanged. A relative energy of -2.96 eV was obtained by subtracting the bare surface energy and ignoring the Al atom[1]. We also prepared additional configurations simulating a feasible onward incorporation and migration pathway, subject to the restricted size of the simulation cell.

Figure 8 shows the MEPs for these reactions. Incorporation (B021-D095) remains metastable but the stability loss is reduced to 0.14 eV from 0.66 eV obtained in the presence of adsorbate H. The first post-incorporation migration step (D095-D096) yields no stability gain but the second (D096-D097) indicates a gain of 0.25 eV, now sufficient to overcome the loss on incorporation. The energy barrier to incorporation is 1.17 eV, corresponding to a PALE process temperature of 400 K for reasonable activation within 60 seconds. The migration steps have lower barriers (respectively 0.65 eV and 0.80 eV) which would be surmountable at this temperature. The effective barrier to migration is lower than that presented by the reversal of incorporation, and so migration is favoured on both kinetic and thermodynamic grounds.

---

[1] VASP calculates binding energies, so the energy due to a lone atom is negligible.

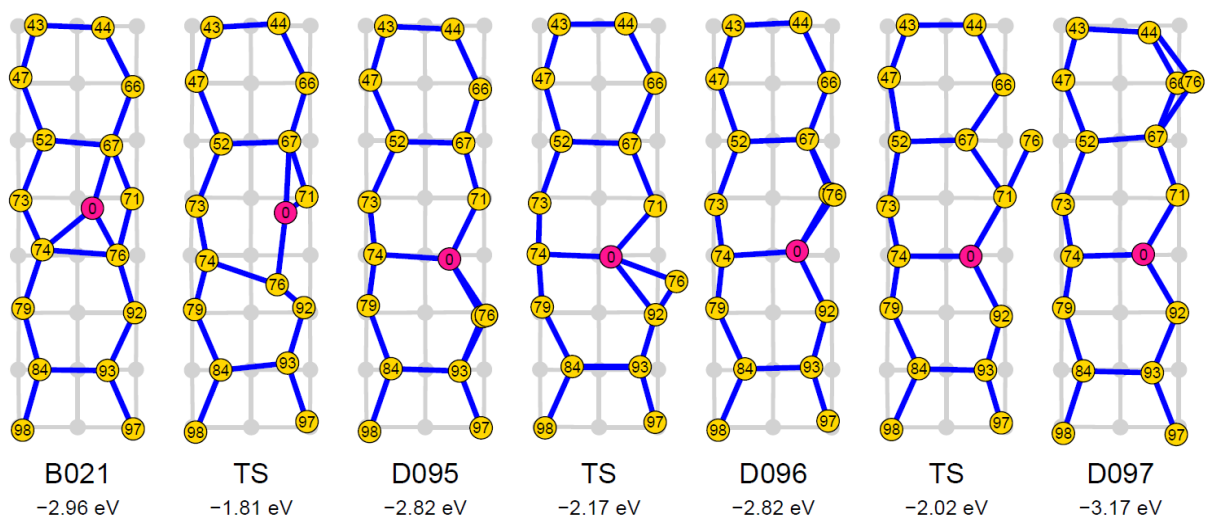

*Figure 8* MEP representation of an Al incorporation and Si migration pathway from the fully-dissociated configuration B021 to a low-energy configuration D097 via metastable configurations D095 and D096. Configuration B021 was derived from B008 by removing its dissociated H. Configurations D095 and D096 are isomeric and a consequence of the restricted cell size. The migrating Si atom is 76. Ultimately, configuration D097 is kinetically and thermodynamically favoured (see text). Energies are relative to the bare surface energy. Image derivation and colouring convention as for figure 3.

## 4 Conclusion

Building on the results of an earlier structural survey [7], we have used climbing-image NEB DFT calculations and transition state theory to analyse the decomposition and incorporation of alane ($AlH_3$) on the H-passivated Si(100) surface. We find that decomposition, resulting in a bridged Al adatom and surface H, proceeds without significant kinetic barrier but, as with P, the energetics of the incorporated Al are close to those of the Al adatom. Furthermore the constraints of the PALE scenario, with reactions confined to an area spanning three or four adjacent dimers, prevent the natural evolution of a stable incorporation configuration. Rather, we would expect Al adatoms to become 'trapped' on the surface at room temperature and incorporate reversibly at elevated temperatures.

However, we describe a scenario in which a bridged Al adatom is destabilized by the removal of dissociated H after decomposition, and incorporation eventually stabilized by surface migration of the ejected Si adatom. The migration pathway is determined by the extent of the surface exposure after decomposition and it is likely that other routes to stable incorporation could be found.

The ability to incorporate acceptor dopants as well as donors in Si(001) with atomic precision will significantly advance the capabilities of patterned ALE. It opens the possibility of p-n junctions fabricated with atomic precision, as well as local control of the electrostatic potential using both positive and negative dopant ions. We keenly anticipate experimental measurements of these structures as a first realisation of this.

# 5 Acknowledgements

The authors acknowledge useful discussions with James Owen and John Randall of Zyvex Labs.

This work acknowledges financial support from the EPSRC ADDRFSS Programme grant (EP/M009564/1).

We are grateful for computational support from the UK national high-performance computing service, ARCHER, for which access was obtained via the UKCP consortium and funded by EPSRC grant ref EP/K013564/1.